# Comprehensive verification of new method "Ethanol as Internal Standard" for determination of volatile compounds in alcohol products by gas chromatography

**Siarhei V. Charapitsa**[1], **Svetlana N. Sytova**[1], **Mikhail G. Markovsky**[2],
**Yurii F. Yakuba**[2], **Yurii N. Kotov**[3]

[1]Research Institute for Nuclear Problems of Belarusian State University
POB 220030, Bobryiskaya Str., 11, Minsk, Belarus,
Tel/Fax: 375-172-265124, e-mail: chere@inp.bsu.by

[2]State Scientific Organization North-Caucasian Zonal Research Institute of Horticulture and Viticulture of the Federal Agency of Scientific Organizations,
40 Let Pobedy Str., 39, Krasnodar 350901, Russia,
Tel/Fax: 7-861-2527074, e-mail: 8612525877@mail.ru

[3]Branch of Joint Stock Company "Rosspirtprom" Wine and Distillery Plant "Cheboksary",
POB 428018, K. Ivanov Str., 63, Cheboksary, Russia,
Tel/Fax: +7-835-2585461, e-mail: kotov.yuri@rosspirtprom.ru

Recently proposed new method "Ethanol as Internal Standard" for determination of volatile compounds in alcohol products by gas chromatography is investigated from different sides. Results of experimental study from three different laboratories from Belarus and Russian Federation are presented.

## 1. INTRODUCTION

The international regulation documents for quality and safety control of alcohol production [1, 2] prescribe determination of the following volatile compounds: acetaldehyde, methyl acetate, ethyl acetate, methanol, 2-propanol, 1-propanol, isobutyl alcohol, n-butanol, isoamyl alcohol. Results of the analysis are expressed in milligrams per litre (mg/L) of absolute alcohol (AA). Such analysis is carried out by the Internal Standard (IS) method. 1-pentanol and 2-pentanol are most commonly used as IS. This method ensures high data reliability. However, the procedure of introducing of an internal standard substance in the sample at the level of some ppm requires a high level of laboratory technicians and performing analyses. For this reason in some national standards the method of External Standard (ES) is used [3, 4]. Finally, to obtain quantitative values of analyzed volatile compounds in mg per litre of absolute alcohol, it is necessary to measure alcohol strength by volume (% v/v) of the analysed sample [1-4].

It was proposed [5] to use ethanol as IS for the analysis of alcohol products in order to increase the accuracy of measurements and obviate the need for the IS addition. The concentration of ethanol in this production can vary from 15% to 96%. The concentration of volatile compounds lies within the range from ppm in rectified alcohol to 30% for the intermediate alcohol products. As a result, the signals from ethanol and from impurities should be registered in a linear range [6]. Nowadays, the testing laboratories are equipped with modern instrumentation for the analysis of alcohol-containing products. These current-technology gas chromatographs have a linear range of registration of seven orders of magnitude that fully obeys the above requirement. Analysis of alcohol products in this case consists in the traditional procedure of determining the relative ratios of the detector response (Relative Response Factors - RRF) of analysed compounds with respect to ethanol by standard solutions and then the subsequent use of these coefficients in the calculation of concentration of analysed volatile compounds. It should be noted that for modern chromatographs coefficients RRF are enough stable and can be tabulated [7].

## 2. MATERIALS AND METHODS

To continue comprehensive examination of the proposed method there were planned and carried out experimental studies with different gas chromatographs (GC) in three different test laboratories: Laboratory of analytical research (LAR) of Research Institute for Nuclear Problems of Belarusian State University (Minsk, Belarus), GC Crystal 5000 (JSC SDO "Chromatec", Russia); in the Scientific centre "Vinodelie" (SCV) of North-Caucasian Zonal Research Institute of Horticulture and Viticulture (Krasnodar, Russia), GC Crystal 2000M (JSC SDO "Chromatec", Russia); control laboratory (CL) of Branch of Joint Stock Company "Rosspirtprom" Wine and Distillery Plant "Cheboksary" (Cheboksary, Russia), GC HP6890 (Agilent Technologies, USA). All mentioned GC were equipped with flame ionization detector (FID). All individual standard compounds were purchased from Sigma-Fluka-Aldrich (Germany). The standard solutions were prepared by adding the individual standard compounds to the ethanol-water mixture (96:4) by gravimetric method according to ASTM D 4307 recommendations [8]. Concentrations of volatile compounds in prepared standard solutions were calculated according to the official method of measurement No. 253.0169/01.00258/2013, certified by Federal Agency for Technical Regulation and Metrology (Rosstandart).

## 3. RESULTS AND DISCUSSION

The first series of experimental research has been performed in SCV on the GC Crystal 2000M. Content of volatile compounds in the first series of the prepared standard solutions was chosen like in cognac and brandy products. Cyclohexanol was added as IS. Concentrations of volatile compounds in the solutions prepared by gravimetric method and calculated concentrations on the base of measured raw data are given in Table 1. The analysis of the experimental data presented in Table 1 shows that the value of relative bias in the determination of the volatile compound concentrations in experiments in the whole range of concentrations for all fifteenth examined compounds does not exceed 10%.

For illustrative purposes the experimental data are presented in Figure 1 and Figure 2 for the following main analyzed components: acetaldehyde, methyl acetate, ethyl acetate, methanol, 2-propanol, 1-propanol, isobutyl alcohol, isoamyl acetate, 1-butanol, isoamyl alcohol, ethyl hexanoate, ethyl octanoate, ethyl decanoate, benzyl alcohol and 2-phenylethanol. The presented graphs show the linear dependence of the detector response (triangle marked) and concentration (circle marked) in mg/L (AA) on the amount of the examined component coming directly to the detector.

Table 1. The experimental data from CSV. Comparison of experimentally measured concentrations of analyzed volatile compounds in standard solutions obtained by tree methods: cyclohexanol as IS, the ES method and method of using ethanol as IS with initial concentration according to gravimetric method.

| Sample | | acetaldehyde | methyl acetate | ethyl acetate | methanol | 2-propanol | ethanol | 1-propanol | isobutyl alcohol | isoamyl acetate | 1-butanol | isoamyl alcohol | ethyl hexanoate | cyclohexanol | ethyl octanoate | ethyl decanoate | benzyl alcohol | 2-phenylethanol |
|---|---|---|---|---|---|---|---|---|---|---|---|---|---|---|---|---|---|---|
| | | | | | | | Concentration, mg /L (AA) | | | | | | | | | | | |
| **A (5000)** | gravimetric method | 5870 | 5642 | 5681 | 5894 | 5619 | 789300 | 5523 | 5556 | 5545 | 5597 | 5657 | 5580 | 111 | 5454 | 5745 | 5642 | 5766 |
| | qunt (0,5 mcl Split=20) | 128,7 | 123,7 | 124,6 | 129,2 | 123,2 | 17307 | 121,1 | 121,8 | 121,6 | 122,7 | 124,1 | 122,4 | 2,4 | 119,6 | 126,0 | 123,7 | 126,4 |
| | response x10, pC | 7456 | 6973 | 9329 | 6304 | 8757 | 616442 | 9811 | 12285 | 8599 | 11547 | 9624 | 10121 | 263 | 10984 | 10529 | 5865 | 6812 |
| | cyclohesanol as IS | 5870 | 5597 | 5679 | 5883 | 5630 | 790426 | 5529 | 5567 | 5612 | 5565 | 5677 | 5603 | 111 | 5470 | 5751 | 5660 | 5780 |
| | repeat, % | 0,5 | 2,1 | 1,9 | 1,5 | 0,9 | 1,9 | 0,6 | 0,9 | 0,6 | 0,5 | 0,4 | 0,3 | 0,0 | 0,2 | 0,6 | 0,3 | 0,4 |
| | relative bias, % | 0,0 | -0,8 | 0,0 | -0,2 | 0,2 | 0,1 | 0,1 | 0,2 | 0,3 | 0,4 | 0,3 | 0,4 | 0,0 | 0,3 | 0,1 | 0,3 | 0,2 |
| | ES | 5694 | 5415 | 5499 | 5692 | 5474 | 657525 | 5362 | 5398 | 5445 | 5397 | 5510 | 5435 | 91 | 5304 | 5577 | 5502 | 5617 |
| | repeat, % | 11,7 | 13,3 | 13,1 | 12,7 | 12,1 | 13,1 | 11,8 | 12,1 | 11,7 | 11,7 | 11,6 | 11,5 | 11,2 | 11,4 | 11,8 | 10,9 | 10,8 |
| | relative bias, % | -3,0 | -4,0 | -3,2 | -3,4 | -2,6 | -16,7 | -2,9 | -2,8 | -2,7 | -2,7 | -2,6 | -2,6 | -17,3 | -2,7 | -2,9 | -2,5 | -2,6 |
| | ethanol as IS | 5878 | 5605 | 5688 | 5891 | 5638 | 789300 | 5537 | 5574 | 5620 | 5573 | 5685 | 5610 | 114 | 5477 | 5758 | 5668 | 5788 |
| | repeat, % | 1,4 | 0,3 | 0,0 | 0,4 | 1,0 | 0,0 | 1,2 | 1,0 | 1,3 | 1,4 | 1,5 | 1,6 | 1,9 | 1,6 | 1,3 | 2,2 | 2,3 |
| | relative bias, % | 0,1 | -0,6 | 0,1 | -0,1 | 0,3 | 0,0 | 0,2 | 0,3 | 0,4 | 0,5 | 0,5 | 0,5 | 3,5 | 0,4 | 0,2 | 0,5 | 0,4 |
| **B (1000)** | gravimetric method | 1094 | 1051 | 1058 | 1119 | 1047 | 789300 | 1029 | 1035 | 1033 | 1043 | 1054 | 1040 | 114 | 1016 | 1070 | 1051 | 1074 |
| | qunt (0,5 mcl Split=20) | 25,8 | 24,8 | 25,0 | 26,4 | 24,7 | 18613 | 24,3 | 24,4 | 24,4 | 24,6 | 24,9 | 24,5 | 2,7 | 24,0 | 25,2 | 24,8 | 25,3 |
| | response x10, pC | 56,8 | 45,3 | 65,5 | 56,5 | 81,2 | 55335 | 97,7 | 117 | 100 | 111 | 118 | 105 | 15,1 | 112 | 119 | 98,8 | 129 |
| | cyclohesanol as IS | 1042 | 994,3 | 1006,9 | 1075,8 | 988,0 | 784869 | 990,2 | 997,9 | 1003 | 992,0 | 1013 | 1005 | 117,4 | 993,6 | 1042 | 1001 | 1021 |
| | repeat, % | 0,4 | 1,4 | 0,3 | 0,2 | 0,1 | 0,0 | 0,1 | 0,1 | 0,5 | 0,1 | 0,3 | 0,0 | 0,0 | 0,0 | 0,0 | 0,4 | 0,2 |
| | relative bias, % | -4,7 | -5,4 | -4,9 | -3,9 | -5,6 | -0,6 | -3,8 | -3,6 | -3,8 | -4,0 | -3,9 | -3,3 | 0,0 | -2,2 | -2,7 | -4,8 | -5,0 |
| | ES | 1157 | 1101 | 1115 | 1191 | 1099 | 746942 | 1099 | 1107 | 1114 | 1101 | 1125 | 1116 | 111 | 1103 | 1156 | 1113 | 1136 |
| | repeat, % | 0,7 | 2,6 | 1,5 | 1,4 | 1,2 | 1,3 | 1,1 | 1,1 | 1,0 | 0,7 | 1,0 | 0,8 | 1,2 | 1,2 | 1,7 | 1,6 | 1,0 |
| | relative bias, % | 5,8 | 4,7 | 5,4 | 6,4 | 5,0 | -5,4 | 6,8 | 7,0 | 6,8 | 6,6 | 6,8 | 7,3 | -5,3 | 8,5 | 8,0 | 5,9 | 5,7 |
| | ethanol as IS | 1051 | 1003 | 1016 | 1085 | 996 | 789300 | 999 | 1006 | 1012 | 1000 | 1022 | 1014 | 122 | 1002 | 1051 | 1009 | 1029 |
| | repeat, % | 0,6 | 1,3 | 0,2 | 0,1 | 0,1 | 0,0 | 0,2 | 0,2 | 0,3 | 0,6 | 0,3 | 0,4 | 0,1 | 0,1 | 0,4 | 0,3 | 0,3 |
| | relative bias, % | -3,9 | -4,6 | -4,0 | -3,0 | -4,8 | 0,0 | -3,0 | -2,8 | -3,0 | -3,2 | -3,1 | -2,5 | 4,2 | -1,4 | -1,8 | -4,0 | -4,2 |
| **C (100)** | gravimetric method | 98,1 | 94,2 | 94,8 | 123,7 | 93,9 | 789300 | 92,2 | 92,7 | 93,4 | 93,4 | 94,4 | 93,2 | 114,8 | 91,0 | 95,9 | 94,2 | 96,2 |
| | qunt (0,5 mcl Split=20) | 2,3 | 2,3 | 2,3 | 3,0 | 2,2 | 18911 | 2,2 | 2,2 | 2,2 | 2,2 | 2,3 | 2,2 | 2,7 | 2,2 | 2,3 | 2,3 | 2,3 |
| | response x10, pC | 5,5 | 4,4 | 6,3 | 7,1 | 7,8 | 61162 | 9,8 | 11,6 | 9,9 | 11,0 | 11,7 | 10,4 | 15,9 | 11,2 | 11,9 | 9,7 | 12,6 |
| | cyclohesanol as IS | 94 | 88,02 | 89,02 | 124,70 | 88,14 | 799910 | 91,54 | 91,52 | 91,42 | 90,85 | 92,51 | 91,88 | 114,80 | 91,34 | 96,49 | 91,31 | 92,60 |
| | repeat, % | 0,1 | 0,6 | 0,8 | 0,9 | 0,0 | 0,4 | 0,6 | 0,9 | 0,6 | 0,7 | 0,4 | 1,0 | 0,0 | 0,3 | 0,0 | 0,9 | 0,9 |
| | relative bias, % | -4,1 | -6,6 | -6,1 | 0,8 | -6,1 | 1,3 | -0,7 | -1,3 | -2,1 | -2,7 | -2,0 | -1,4 | 0,0 | 0,4 | 0,6 | -3,1 | -3,7 |
| | ES | 110 | 102,6 | 103,9 | 145,4 | 101,2 | 801919 | 107,0 | 107,0 | 106,9 | 106,2 | 108,3 | 107,5 | 114,5 | 106,8 | 112,8 | 107,0 | 108,5 |
| | repeat, % | 4,1 | 4,7 | 4,9 | 5,0 | 4,6 | 4,6 | 4,7 | 5,0 | 4,7 | 4,8 | 4,5 | 5,1 | 4,1 | 4,4 | 4,1 | 3,3 | 3,2 |
| | relative bias, % | 12,1 | 8,9 | 9,6 | 17,6 | 7,7 | 1,6 | 16,1 | 15,4 | 14,5 | 13,7 | 14,7 | 15,3 | -0,3 | 17,4 | 17,6 | 13,6 | 12,8 |
| | ethanol as IS | 93 | 87,10 | 88,09 | 123,39 | 87,22 | 789300 | 90,57 | 90,55 | 90,45 | 89,88 | 91,53 | 90,90 | 113,99 | 90,36 | 95,46 | 90,34 | 91,62 |
| | repeat, % | 0,5 | 0,1 | 0,3 | 0,4 | 0,4 | 0,0 | 0,2 | 0,4 | 0,1 | 0,3 | 0,0 | 0,5 | 0,4 | 0,1 | 0,4 | 1,3 | 1,4 |
| | relative bias, % | -5,2 | -7,5 | -7,1 | -0,3 | -7,1 | 0,0 | -1,8 | -2,3 | -3,2 | -3,8 | -3,0 | -2,5 | -0,7 | -0,7 | -0,5 | -4,1 | -4,8 |
| **D (10)** | gravimetric method | 8,06 | 7,65 | 7,71 | 33,66 | 7,72 | 789300 | 7,49 | 7,54 | 7,59 | 7,59 | 7,67 | 7,57 | 93,51 | 7,40 | 7,79 | 7,65 | 7,82 |
| | qunt (0,5 mcl Split=20) | 0,193 | 0,184 | 0,185 | 0,808 | 0,185 | 18938 | 0,180 | 0,181 | 0,182 | 0,182 | 0,184 | 0,182 | 2,244 | 0,178 | 0,187 | 0,184 | 0,188 |
| | response x10, pC | 0,49 | 0,41 | 0,56 | 1,93 | 0,66 | 56909 | 0,85 | 0,99 | 0,87 | 0,94 | 1,02 | 0,89 | 12,73 | 0,95 | 1,00 | 0,85 | 1,12 |
| | cyclohesanol as IS | 8,56 | 8,27 | 8,06 | 34,55 | 7,70 | 765233 | 8,04 | 8,07 | 8,21 | 8,23 | 8,22 | 8,11 | 93,51 | 8,03 | 8,41 | 8,19 | 8,48 |
| | repeat, % | 3,0 | 4,5 | 1,3 | 0,3 | 2,1 | 1,7 | 2,9 | 2,7 | 4,4 | 0,9 | 0,1 | 2,2 | 0,0 | 1,4 | 4,7 | 0,9 | 3,1 |
| | relative bias, % | 2,2 | 3,9 | 0,7 | 1,7 | -4,0 | -3,0 | 3,3 | 3,0 | 4,1 | 4,4 | 3,1 | 3,1 | 0,0 | 4,4 | 3,9 | 3,0 | 4,3 |
| | ES | 10,55 | 9,69 | 9,46 | 40,54 | 9,08 | 773603 | 9,46 | 9,75 | 9,67 | 9,68 | 9,68 | 9,55 | 93,78 | 9,45 | 10,00 | 9,66 | 9,99 |
| | repeat, % | 9,0 | 3,5 | 0,3 | 0,8 | 3,1 | 2,7 | 1,9 | 3,5 | 5,5 | 1,9 | 0,9 | 3,2 | 1,0 | 2,4 | 3,5 | 1,9 | 4,1 |
| | relative bias, % | 25,9 | 21,9 | 18,2 | 19,3 | 13,1 | -2,0 | 21,5 | 24,5 | 22,5 | 22,7 | 21,3 | 21,3 | 0,3 | 22,9 | 23,5 | 21,4 | 22,9 |
| | ethanol as IS | 8,83 | 8,53 | 8,32 | 35,66 | 7,94 | 789300 | 8,30 | 8,44 | 8,47 | 8,50 | 8,48 | 8,37 | 95,69 | 8,28 | 8,68 | 8,45 | 8,74 |
| | repeat, % | 1,3 | 6,2 | 3,0 | 1,9 | 0,4 | 0,0 | 4,5 | 3,6 | 2,8 | 0,8 | 1,8 | 0,5 | 1,7 | 0,3 | 3,1 | 0,7 | 1,4 |
| | relative bias, % | 5,4 | 7,3 | 3,9 | 5,0 | -1,0 | 0,0 | 6,6 | 7,7 | 7,4 | 7,7 | 6,3 | 6,4 | 2,3 | 7,7 | 7,1 | 6,2 | 7,6 |
| **E (2)** | gravimetric method | 1,871 | 1,702 | 1,714 | 27,471 | 1,795 | 789300 | 1,666 | 1,676 | 1,689 | 1,689 | 1,707 | 1,684 | 92,139 | 1,646 | 1,733 | 1,702 | 1,740 |
| | qunt (0,5 mcl Split=20) | 0,045 | 0,041 | 0,041 | 0,659 | 0,043 | 18938 | 0,040 | 0,040 | 0,041 | 0,041 | 0,041 | 0,040 | 2,211 | 0,039 | 0,042 | 0,041 | 0,042 |
| | response x10, pC | 0,104 | 0,083 | 0,107 | 1,476 | 0,135 | 55264 | 0,156 | 0,194 | 0,171 | 0,182 | 0,182 | 0,167 | 11,433 | 0,180 | 0,216 | 0,164 | 0,228 |
| | cyclohesanol as IS | 2,00 | 1,80 | 1,74 | 28,3 | 1,77 | 802125 | 1,69 | 1,68 | 1,71 | 1,74 | 1,68 | 1,72 | 92,10 | 1,61 | 1,89 | 1,72 | 1,86 |
| | repeat, % | 2,5 | 9,0 | 4,0 | 5,2 | 1,6 | 6,5 | 4,4 | 1,1 | 2,8 | 7,5 | 7,3 | 0,0 | 0,0 | 4,6 | 6,1 | 0,6 | 0,2 |
| | relative bias, % | 3,3 | 1,7 | -2,4 | 2,9 | -5,0 | 1,6 | -2,6 | -3,8 | -2,3 | -0,8 | -5,5 | -1,8 | 0,0 | -5,6 | 4,9 | -2,6 | 2,9 |
| | ES | 2,31 | 1,95 | 1,76 | 30,1 | 1,61 | 734643 | 1,62 | 1,79 | 1,83 | 1,86 | 1,79 | 1,83 | 83,69 | 1,83 | 2,01 | 1,84 | 1,99 |
| | repeat, % | 1,3 | 12,7 | 1,1 | 5,4 | 1,1 | 1,8 | 13,1 | 4,6 | 0,9 | 13,1 | 7,3 | 7,1 | 0,2 | 16,2 | 6,3 | 0,4 | 0,0 |
| | relative bias, % | 19,3 | 10,1 | -1,3 | 9,3 | -13,7 | -6,9 | -6,6 | 2,5 | 4,1 | 5,9 | 0,8 | 4,6 | -9,1 | 6,9 | 11,8 | 4,1 | 9,9 |
| | ethanol as IS | 1,97 | 1,77 | 1,71 | 27,9 | 1,74 | 789300 | 1,70 | 1,65 | 1,69 | 1,71 | 1,65 | 1,69 | 89,91 | 1,59 | 1,86 | 1,70 | 1,83 |
| | repeat, % | 4,1 | 7,5 | 5,5 | 3,6 | 8,8 | 0,0 | 12,3 | 2,9 | 2,7 | 1,3 | 9,1 | 8,9 | 1,6 | 2,9 | 4,5 | 2,2 | 1,7 |
| | relative bias, % | 1,7 | 0,1 | -3,9 | 1,2 | -6,5 | 0,0 | -1,9 | -5,3 | -3,9 | -2,3 | -6,9 | -3,3 | -2,4 | -7,1 | 3,3 | -4,1 | 1,3 |

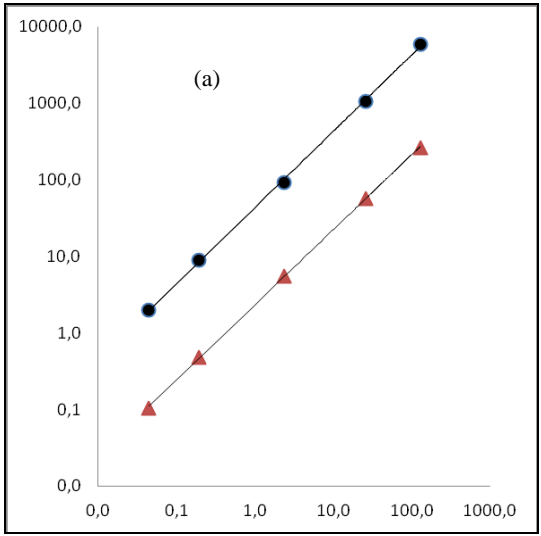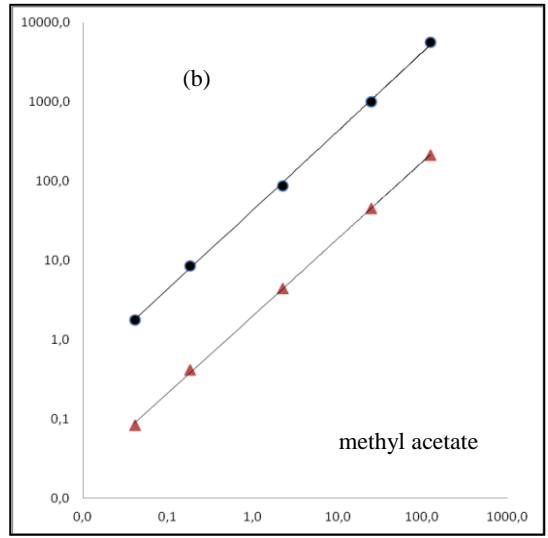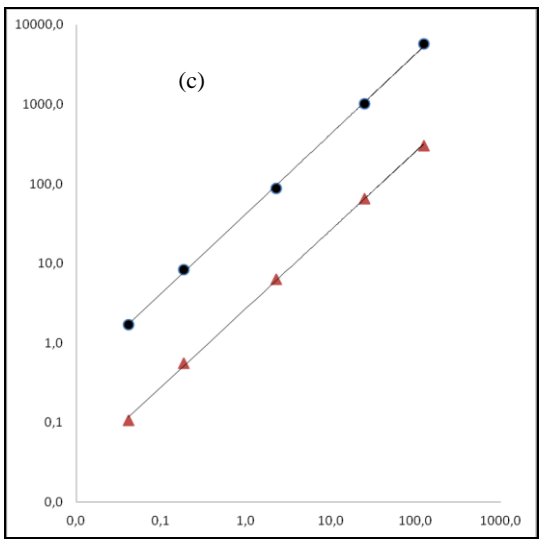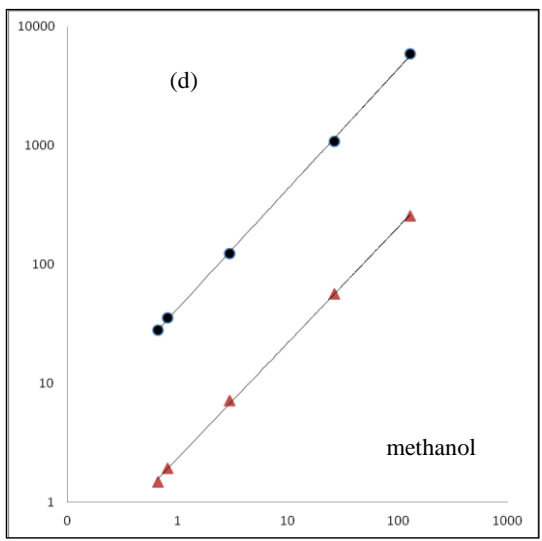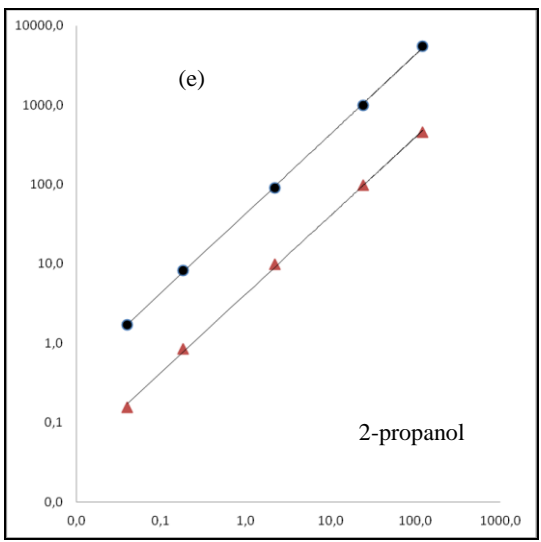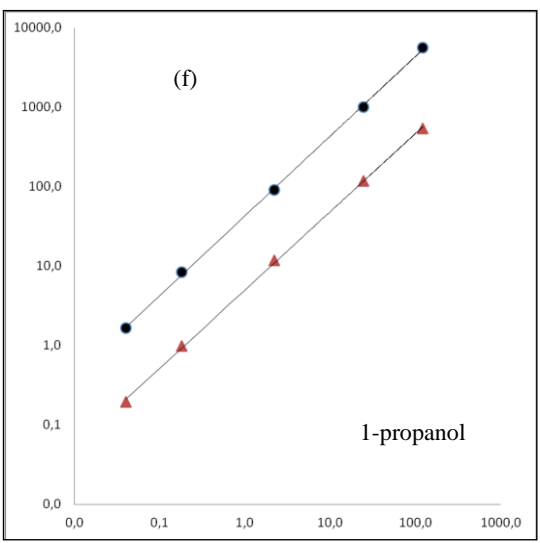

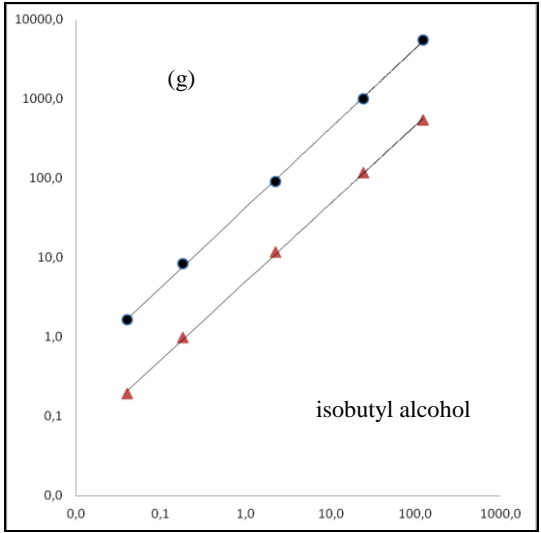

(g) isobutyl alcohol

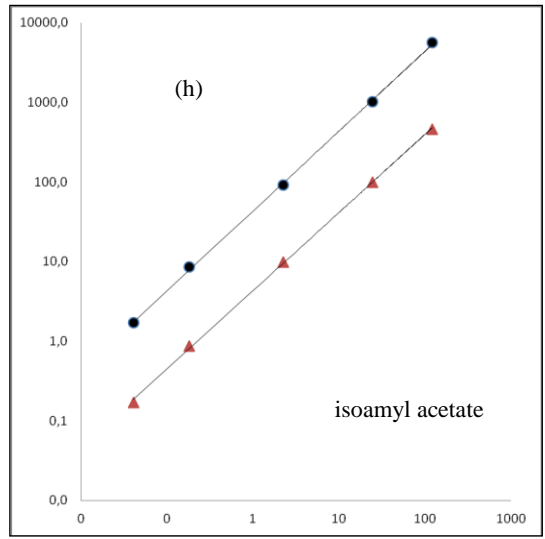

(h) isoamyl acetate

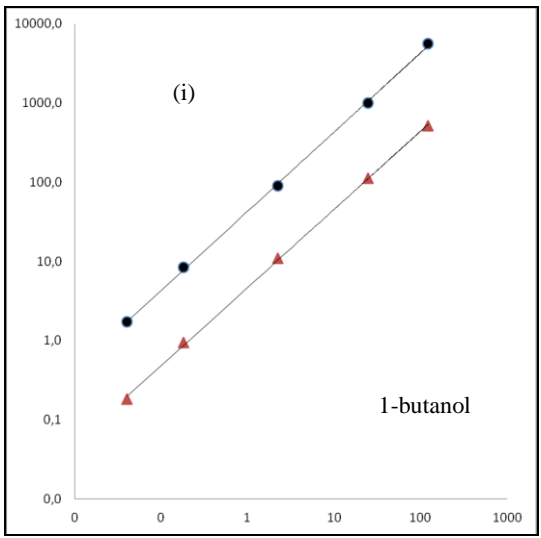

(i) 1-butanol

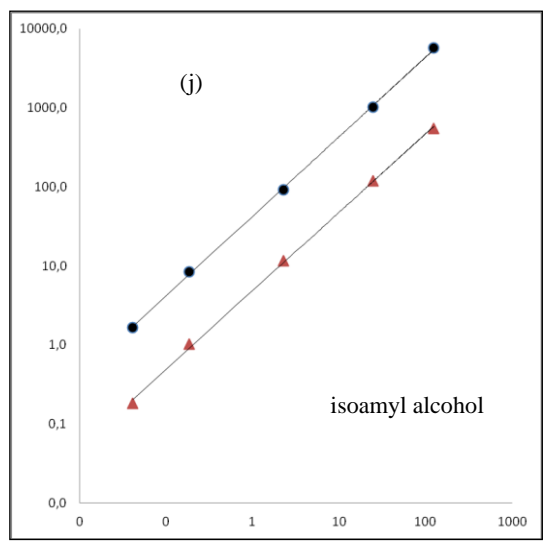

(j) isoamyl alcohol

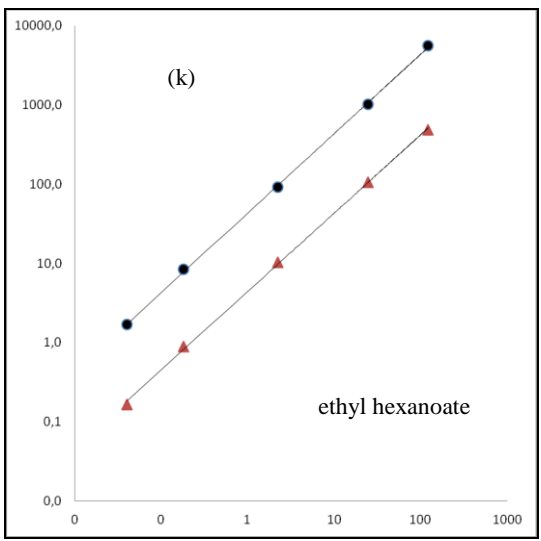

(k) ethyl hexanoate

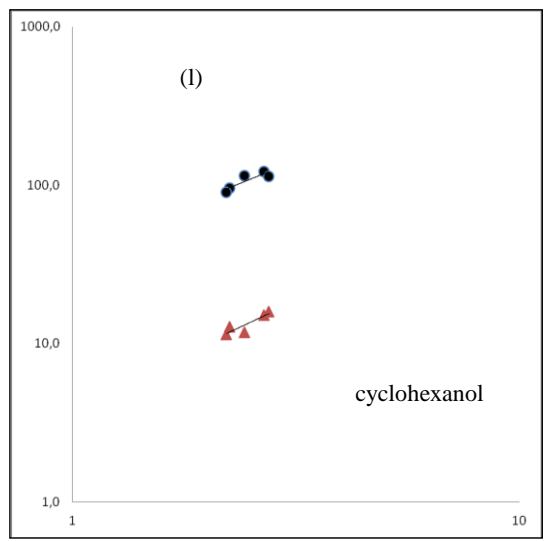

(l) cyclohexanol

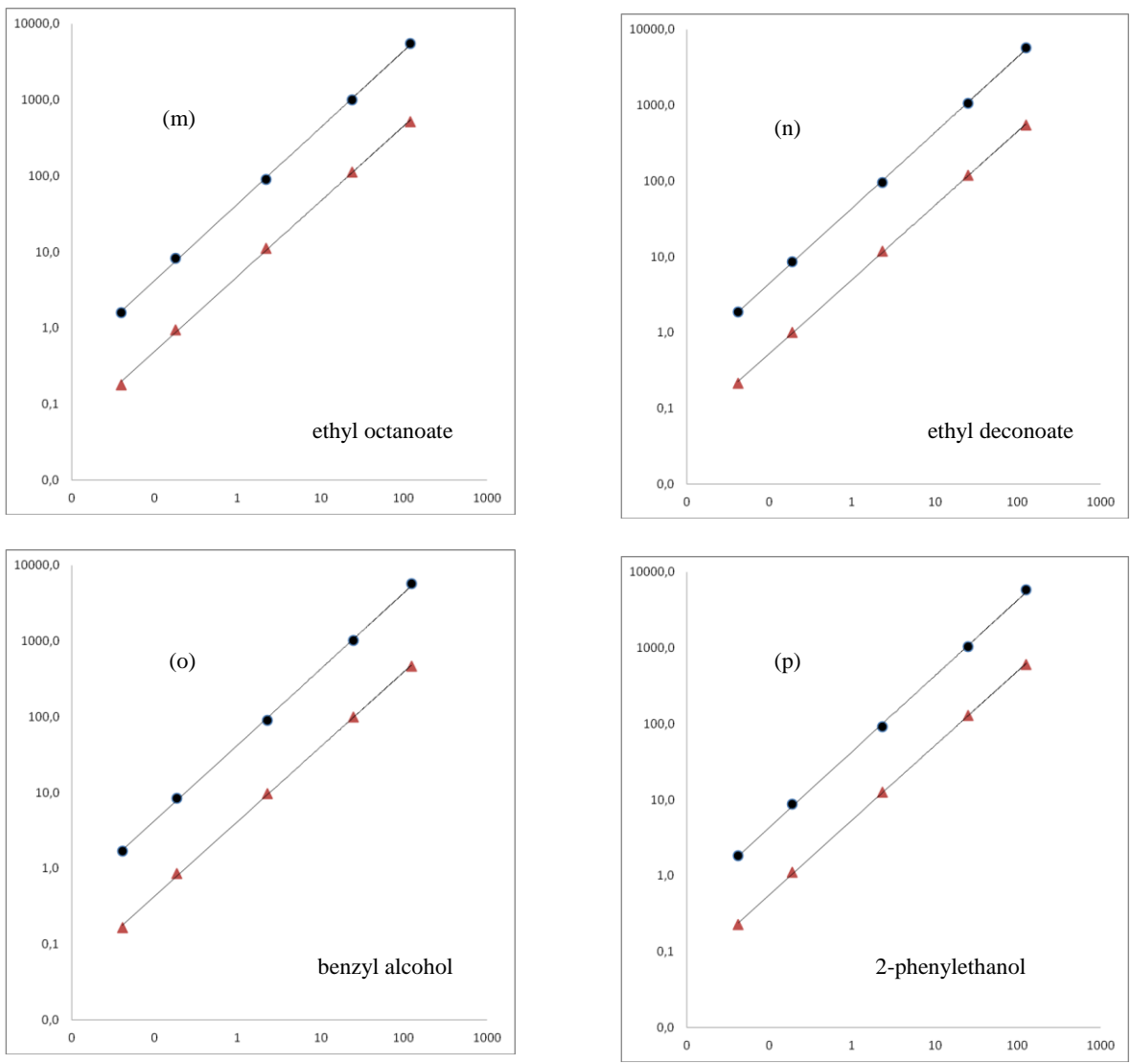

● — concentration, mg/L (AA), ▲ — response × 10, pC, horizontal axis — amount, pg

Figure 1. Experimental results on demonstration of the following compounds: (a) – acetaldehyde, (b) – methyl acetate, (c) – ethyl acetate, (d) – methanol, (e) – 2-propanol, (f) – 1-propanol, (g) – isobutyl alcohol, (h) – isoamyl acetate, (i) – 1-butanol, (j) – isoamyl alcohol, (k) – ethyl hexanoate, (l) – cyclohexanol, (m) – ethyl octanoate, (n) – ethyl decanoate, (o) – benzyl alcohol and (p) – 2-phenylethanol.

The analysis of the experimental data shows that the relative bias between the experimentally measured concentrations calculated in accordance with proposed method using ethanol as IS and the values of concentrations assigned during the preparation by gravimetric method for all analyzed fifteen components in the five analyzed solutions does not exceed 7,7 %. At the same time the relative bias between measured concentrations calculated in accordance with traditional IS method using cyclohexanol as IS and traditional ES method the values of concentrations assigned during the preparation by gravimetric method for all analyzed fifteen components in the five analyzed solutions does not exceed 6,6 % and 25,9 %, respectively.

The second series of experimental research has been performed in LAR and in CL on the GC Crystal 5000 and HP6890, respectively. To demonstrate the reliability of the proposed method the standard ethanol-water (96:4) solution with initial volatile compounds concentration about 4000 mg/L (AA) was analyzed after dilution with water in the ratios 1:1, 1:9, 1:99, 1:1999 and 1:9999.

Experimental results are presented in Table 2 and Table 3. Illustrative presentation of obtained experimental data are in the Figure 3–6.

Even after dilution with water in the ratio 1:999, the difference between the measured concentrations of all compounds and their values calculated using the gravimetric method does not exceed 7.8 %. With the dilution 1:9999 there are peaks of methanol and ethanol only. Other compounds are significantly less than the level of detection. But even in this case the relative discrepancy of measured concentrations of methanol does not exceed 6.6%.

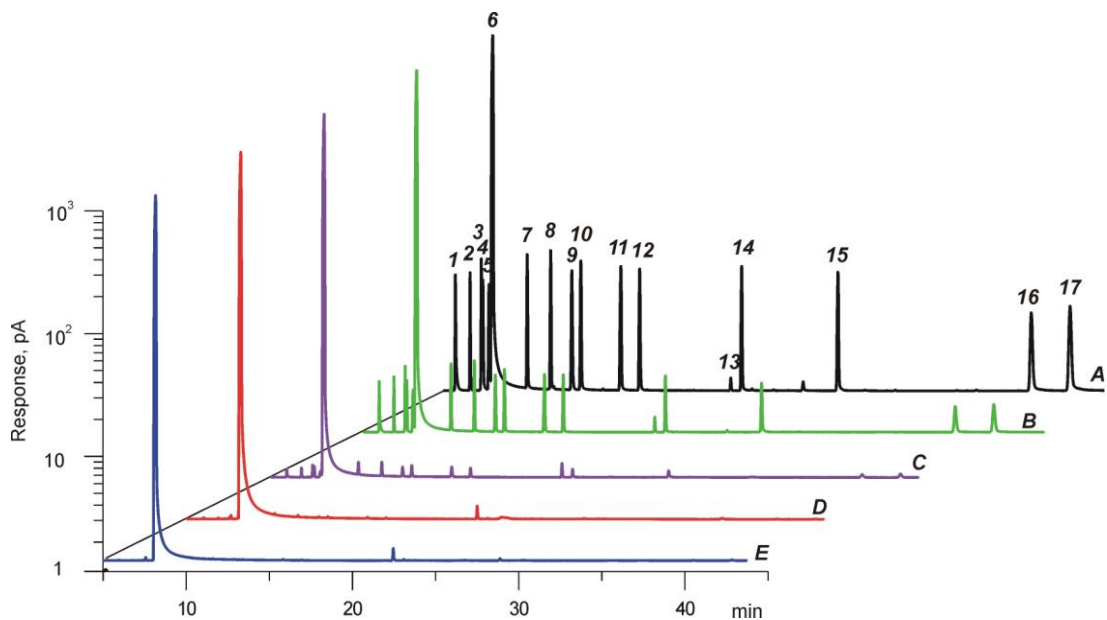

Figure. 2. Chromatograms of standard solutions A-E from Table 1.  1 – acetaldehyde, 2 – methyl acetate, 3 – ethyl acetate, 4 – methanol, 5 – 2-propanol, 6 – ethanol, 7 – 1-propanol, 8 – isobutyl alcohol, 9 – isoamyl acetate, 10 – 1-butanol, 11 – isoamyl alcohol, 12 – ethyl hexanoate, 13 – cyclohexanol, 14 – ethyl octanoate, 15 – ethyl decanoate, 16 – benzyl alcohol, 17 – 2-phenylethanol.

Table 2. The experimental data from LAR. The measured concentrations of analyzed volatile compounds and ethanol, presented according to the degree of dilution with water.

| sample (dilution) | measured concentration mg /L (AA) (relative bias,%) [concentration under certificate mg /L (AA) / mg /L (sol)] {response x10, pC} amount, pg compound | | | | | | | | | |
|---|---|---|---|---|---|---|---|---|---|---|
| | acetaldehyde | methyl acetate | ethyl acetate | methanol | 2-propanol | ethanol | 1-propanol | isobutyl alcohol | 1-butanol | isoamyl alcohol |
| A (No) | 4556 (6,6) [4275/3768] {10720} 91899 | 4436 (0,9) [4397/3875] {11276} 94524 | 4253 (1,9) [4173/3678] {14420} 89710 | 42586 (1,4) [41995/37017] {115328} 902864 | 4112 (3,0) [3991/3518] {16655} 85806 | N/A N/A [789300/695748] {2825852} 16969460 | 4076 (1,6) [4012/3 536] {19676} 86253 | 4049 (1,9) [3975/3504] {22784} 85466 | 4174 (2,5) [4071/3588] {21330} 87522 | 4458 (9,5) [4071/3588] {23143} 87522 |
| B (1:1) | 4451 (4,1) [4275/1884] {4732,6} 43761 | 4127 (-6,1) [4397/1938] {4741} 45012 | 4018 (-3,7) [4173/1839] {6157} 42719 | 40462 (-3,7) [41995/18509] {49525} 429935 | 4000 (0,2) [3991/1759] {7323} 40860 | N/A N/A [789300/347874] {1277251} 8080695 | 3973 (-1,0) [4012/1768] {8668} 41073 | 4007 (0,8) [3975/1752] {10190} 40698 | 4096 (0,6) [4071/1794] {9462} 41677 | 4412 (8,4) [4071/1794] {10353} 41677 |
| C (1:9) | 4340 (1,5) [4275/377] {931,6} 9190 | 3961 (-9,9) [4397/388] {918,7} 9452 | 3780 (-9,4) [4173/368] {1169} 8971 | 39043 (-7,0) [41995/3702] {9647} 90286 | 3875 (-2,9) [3991/352] {1432} 8581 | N/A N/A [789300/69575] {257842} 1696946 | 3868 (-3,6) [4012/354] {1704} 8625 | 3904 (-1,8) [3975/350] {2004} 8547 | 4012 (-1,4) [4071/359] {1870} 8752 | 4318 (6,1) [4071/359] {2045} 8752 |
| D (1:99) | 4406 (3,1) [4275/37,7] {78,57} 919 | 4002 (-9,0) [4397/38,8] {77,10} 945 | 3762 (-9,8) [4173/36,8] {96,66} 897 | 38645 (-8,0) [41995/370,2] {793,5} 9029 | 3866 (-3,1) [3991/35,2] {118,8} 858 | N/A N/A [789300/6958] {21427} 169695 | 3862 (-3,7) [4012/35,4] {141,33} 863 | 3903 (-1,8) [3975/35,0] {166,6} 855 | 4107 (0,9) [4071/35,9] {159,1} 875 | 4479 (10,0) [4071/35,9] {171,9} 875 |
| E (1:999) | 4280 (0,1) [4275/3,77] {7,74} 91,9 | 4292 (-2,4) [4397/3,88] {8,60} 94,5 | 4107 (-1,6) [4173/3,68] {10,7} 89,7 | 38764 (-7,7) [41995/37,02] {80,7} 903 | 3818 (-4,3) [3991/3,52] {12,4} 85,8 | N/A N/A [789300/696] {2173} 16969 | 3820 (-4,8) [4012/3,54] {14,2} 86,3 | 4140 (4,1) [3975/3,50] {17,9} 85,5 | 4024 (-1,2) [4071/3,59] {15,8} 87,5 | 3937 (-3,3) [4071/3,59] {15,8} 87,5 |
| F (1:9999) | N/A | N/A | N/A | 39210 (-6,6) [41995/3,702] {8,12} 90,3 | N/A | N/A N/A [789300/69,6] {223} 1697 | N/A | N/A | N/A | N/A |

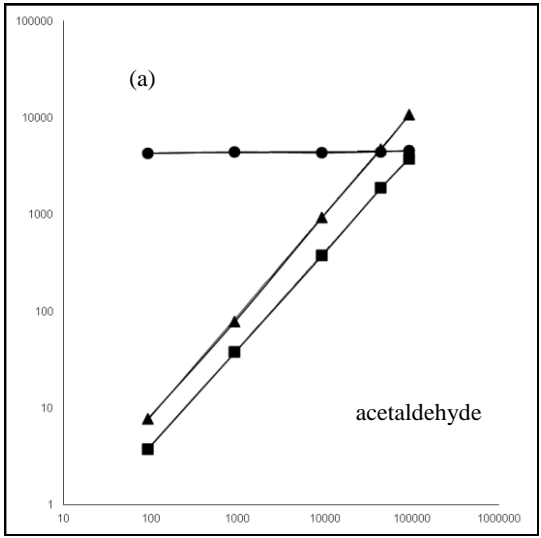
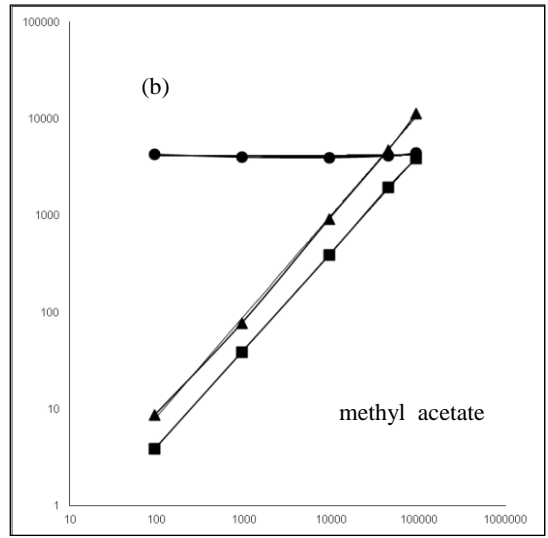
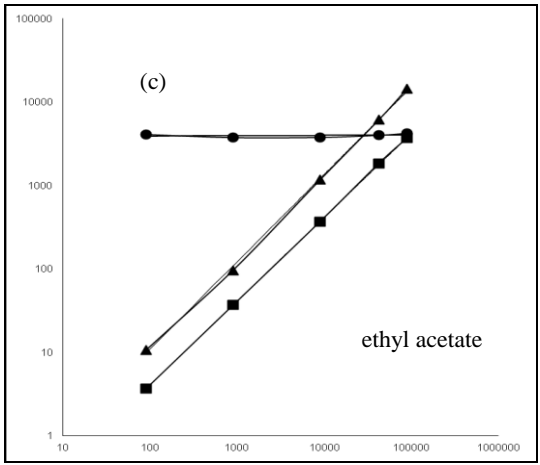
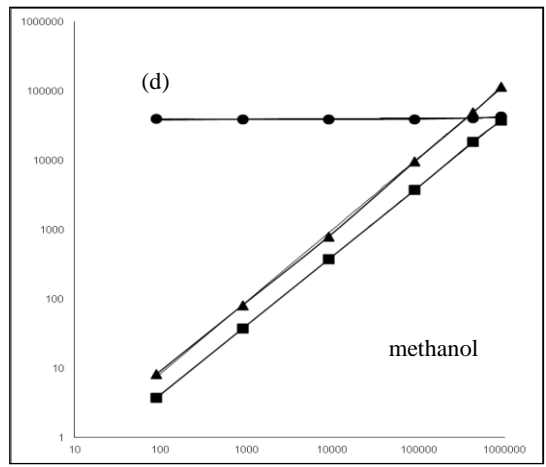
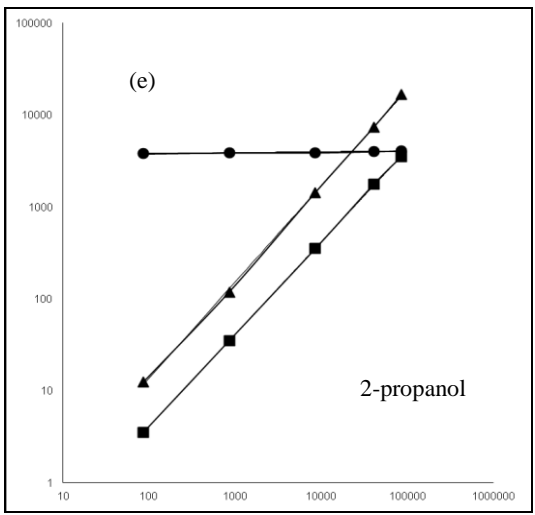
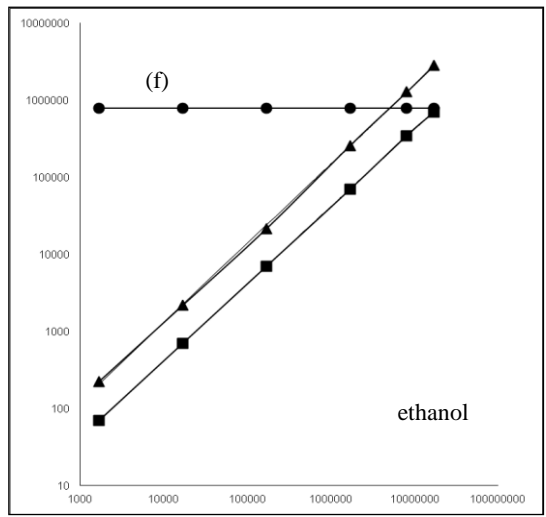

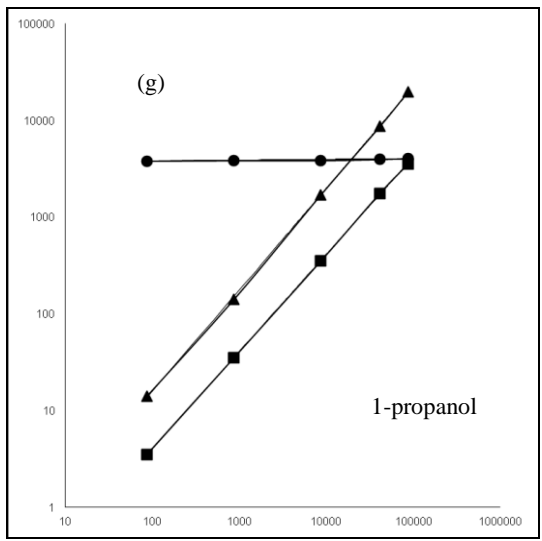
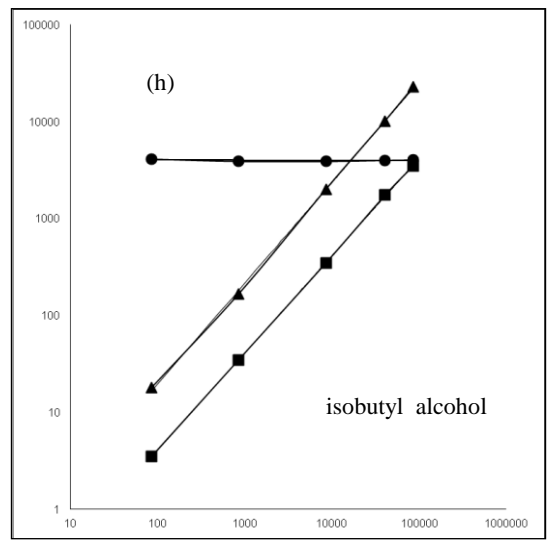
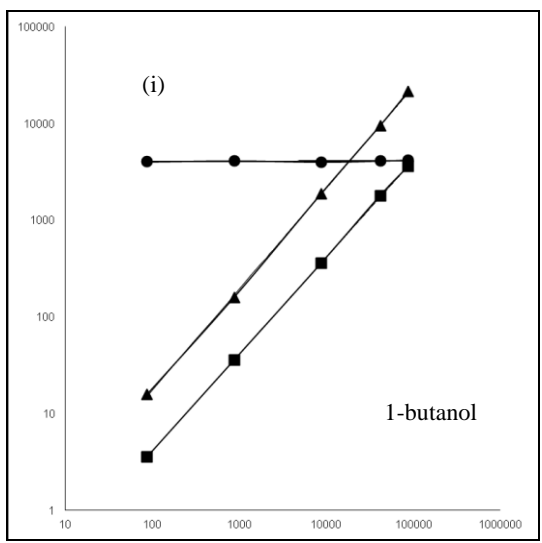
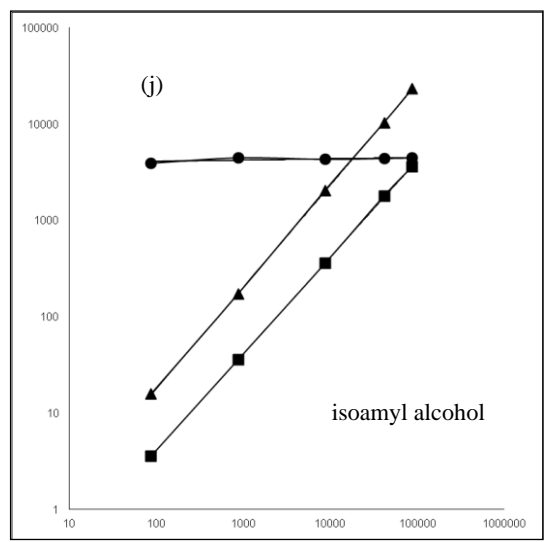

● − concentration, mg/L (AA)   ■ − concentration, mg/L (sol)   ▲ − response x 10, pC,  horizontal axis  − amount, pg

Figure 3. Experimental results on demonstration of the following compounds: (a) – acetaldehyde, (b) – methyl acetate, (c) – ethyl acetate, (d) – methanol, (e) – 2-propanol, (f) – 1-propanol, (g) – isobutyl alcohol, (h) – 1-butanol, (j) – isoamyl alcohol. The first line (circle marked) is concentration of the analysed compound expressed in mg per litre of absolute alcohol. The second line (triangle marked) and the third ones (square marked) are the detector response versus the amount of the compound and the concentration in mg per litre of solution, respectively.

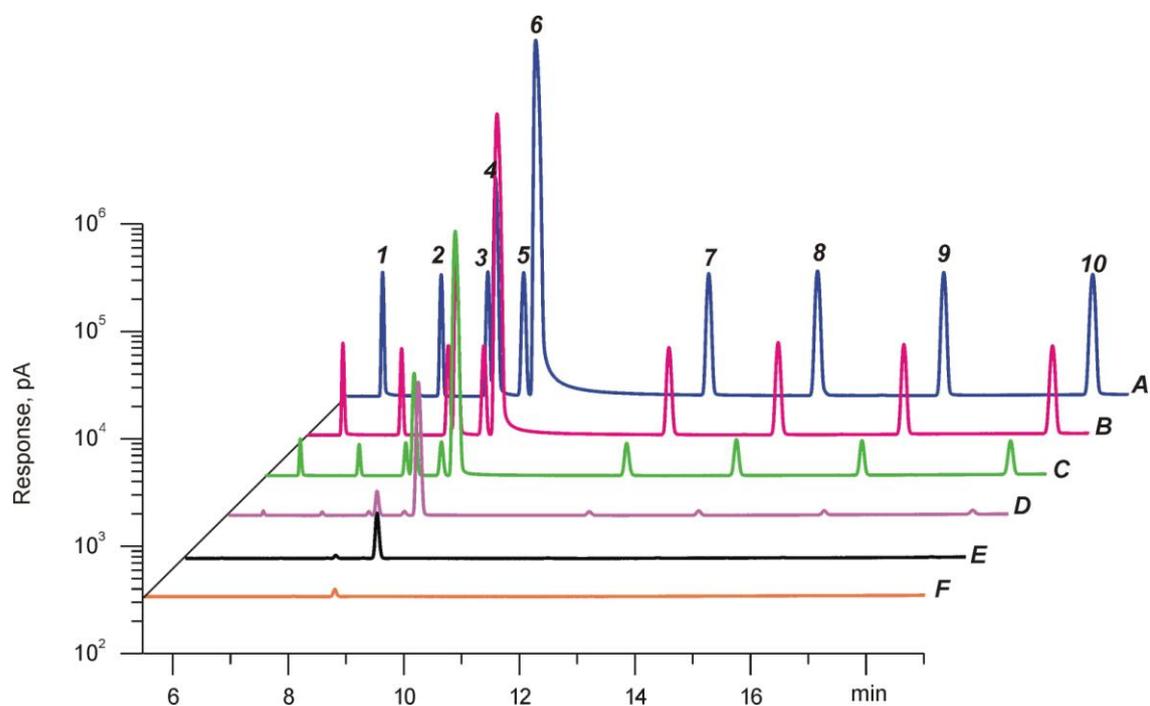

Figure 4. Chromatograms of standard solutions A-F from Table 2.  1 – acetaldehyde, 2 – methyl acetate, 3 – ethyl acetate, 4 – methanol, 5 – 2-propanol, 6 – ethanol, 7 – 1-propanol, 8 – isobutyl alcohol, 9 – n-butanol, 10 - isoamyl alcohol

Table 3. The experimental data from CL. The measured concentrations of analyzed volatile compounds and ethanol, presented according to the degree of dilution with water.

| sample (dilution) | | acetaldehyde | methyl acetate | methanol | 2-propanol | ethanol | 1-propanol | isobutyl alcohol | n-butanol |
|---|---|---|---|---|---|---|---|---|---|
| | compound concentration under certificate, mg /L (AA) | | | | | | | | |
| | concentration under certificate, mg /L (sol) | | | | | | | | |
| | measured concentration, mg /L (AA) | | | | | | | | |
| | relative bias,% | | | | | | | | |
| | amount, pg | | | | | | | | |
| | response x10, pC | | | | | | | | |
| A (No) | | 5170 | 5200 | 5291 | 5242 | 789300 | 7196 | 1171 | 5324 |
| | | 4756 | 4784 | 4868 | 4823 | 726156 | 6620 | 1077 | 4898 |
| | | 5405 | 5127 | 4840 | 5091 | 789300 | 7157 | 1157 | 5358 |
| | | 4,5 | -1,4 | -8,5 | -2,9 | N/A | -0,5 | -1,2 | 0,6 |
| | | 119 | 120 | 122 | 121 | 18154 | 166 | 27 | 122 |
| | | 262640 | 473048 | 551182 | 785344 | 117394309 | 1038326 | 289264 | 1635733,8 |
| B (1:1) | | 5170 | 5200 | 5291 | 5242 | 789300 | 7196 | 1171 | 5324 |
| | | 2378 | 2392 | 2434 | 2412 | 363078 | 3310 | 539 | 2449 |
| | | 5231 | 5292 | 5288 | 5243 | 789300 | 7187 | 1174 | 5311 |
| | | 1,2 | 1,8 | -0,1 | 0,0 | N/A | -0,1 | 0,2 | -0,3 |
| | | 59,5 | 59,8 | 60,8 | 60,3 | 9077,0 | 82,8 | 13,5 | 61,2 |
| | | 131420,7 | 256031 | 309192 | 422395 | 60259800 | 534562 | 150959 | 831805 |
| C (1:9) | | 5170 | 5200 | 5291 | 5242 | 789300 | 7196 | 1171 | 5324 |
| | | 476 | 478 | 487 | 482 | 72616 | 662 | 108 | 490 |
| | | 3936 | 6849 | 5107 | 9092 | 789300 | 5219 | 1216 | 6861 |
| | | -1,0 | -1,8 | 0,8 | -1,3 | N/A | -0,5 | -1,5 | -0,8 |
| | | 11,9 | 12,0 | 12,2 | 12,1 | 1815,4 | 16,6 | 2,7 | 12,2 |
| | | 28419 | 55123 | 70780 | 89072 | 12827790 | 116146 | 31552 | 176935 |
| D (1:99) | | 5170 | 5200 | 5291 | 5242 | 789300 | 7196 | 1171 | 5324 |
| | | 47,6 | 47,8 | 48,7 | 48,2 | 7261,6 | 66,2 | 10,8 | 49,0 |
| | | 3815 | 6781 | 4919 | 8321 | 789300 | 5075 | 1167 | 6648 |
| | | 0,5 | 1,3 | 0,0 | -0,2 | N/A | 0,1 | 0,0 | -0,2 |
| | | 1,19 | 1,20 | 1,22 | 1,21 | 181,54 | 1,66 | 0,27 | 1,22 |
| | | 3267 | 6580 | 7903 | 9631 | 1499539 | 13284 | 3594 | 20158 |
| E (1:999) | | 5170 | 5200 | 5291 | 5242 | 789300 | 7196 | 1171 | 5324 |
| | | 4,76 | 4,79 | 4,87 | 4,83 | 726,88 | 6,63 | 1,08 | 4,90 |
| | | 4474 | 5388 | 5233 | 4717 | 789300 | 6553 | 1090 | 4803 |
| | | -13,5 | 3,6 | -1,1 | -10,0 | N/A | -8,9 | -7,0 | -9,8 |
| | | 0,119 | 0,120 | 0,122 | 0,121 | 18,172 | 0,166 | 0,027 | 0,123 |
| | | 0,119 | 0,120 | 0,122 | 0,121 | 18,172 | 0,166 | 0,027 | 0,123 |

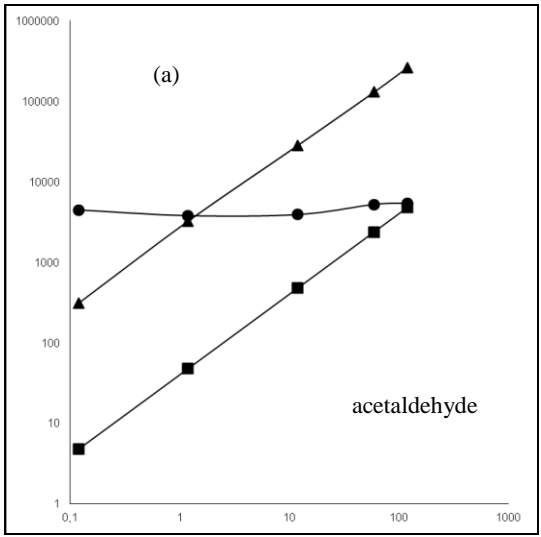
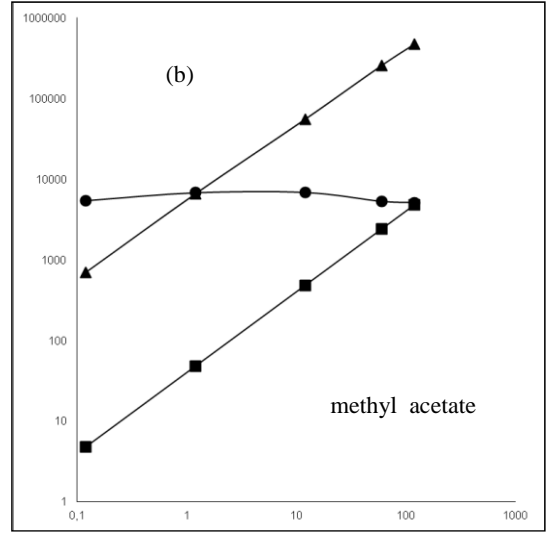
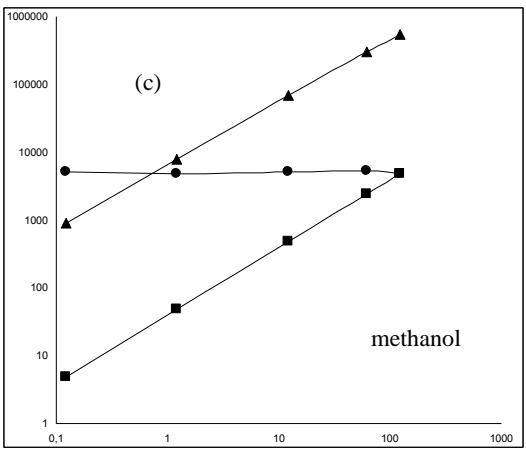
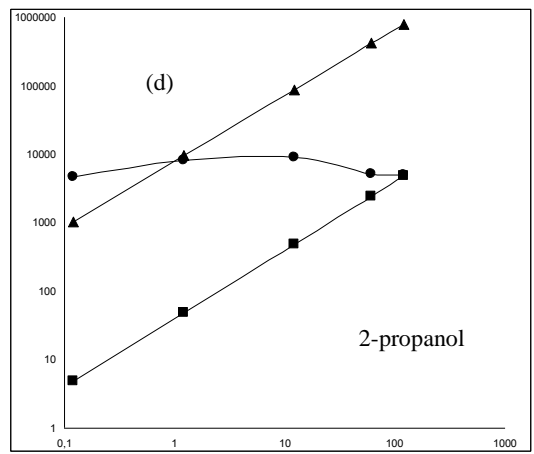
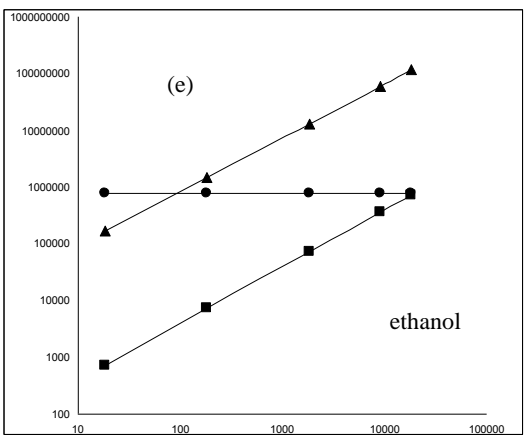
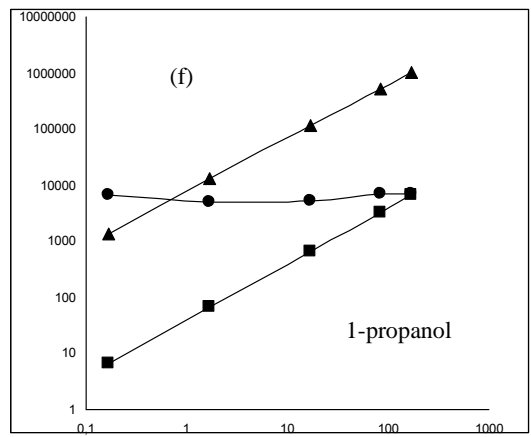

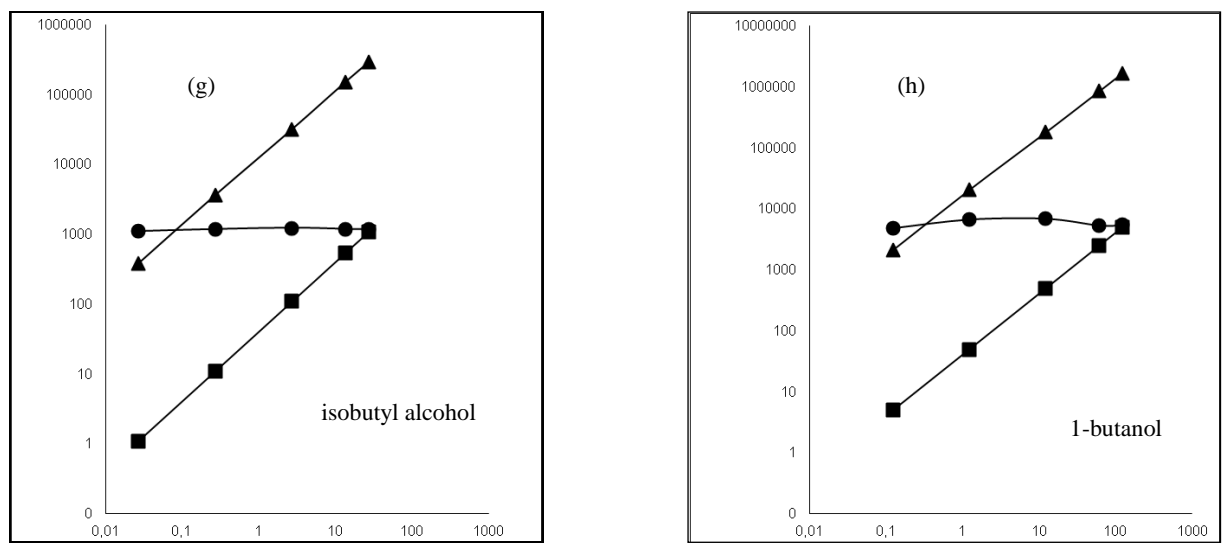

● − concentration, mg/L (AA)  ■ − concentration, mg/L (sol)  ▲ − response x 10, pC,  horizontal axis − amount, pg

Figure 5. Experimental results on demonstration of the following compounds: (a) – acetaldehyde, (b) – methyl acetate, (c) – methanol, (d) – 2-propanol, (e) – ethanol, (f) – 1-propanol, (g) – isobutyl alcohol, (h) – isoamyl acetate, (i) – 1-butanol. (The first line (circle marked) is concentration of the analysed compound expressed in mg per litre of absolute alcohol. The second line (triangle marked) and the third ones (square marked) are the detector response versus the amount of the compound and the concentration in mg per litre of solution, respectively.

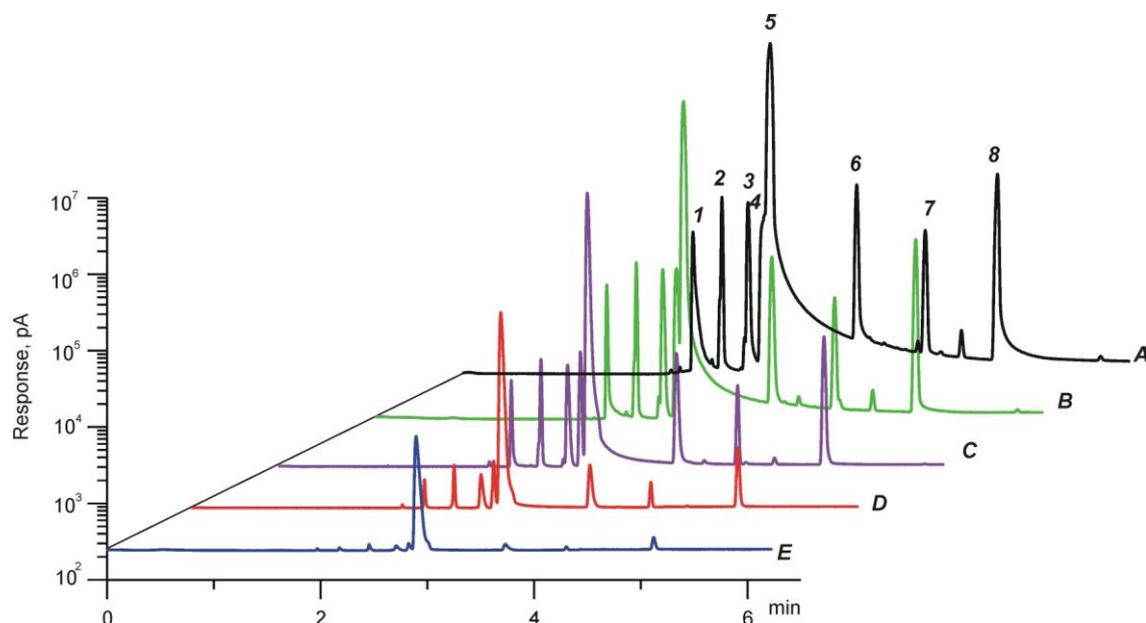

Figure 6. Chromatograms of standard solutions A-F from Table 3.  1 − acetaldehyde, 2 − methyl acetate, 3 − methanol, 4 − 2-propanol, 5 − ethanol, 6 − 1-propanol, 7 − isobutyl alcohol, 8 − 1-butanol.

It is important to note that at present time practically all manufactures produce GC with a wide linear dynamic range of FID and with high stable technical parameters. For illustration in the Table 4 are presented experimental coefficients RRF from different 13 laboratories. Coefficients RF for ethanol are presented in the bottom line. These coefficients RRF and RF were obtained during adjustment works of new GC in control laboratories of distillery plants.

Table 4. Values of coefficients $RRF_i^{Et}$ and RF for ethanol from 13 different laboratories.

| compound | RRF (relative to ethanol) | | | | | | | | | | | | | RRF average | RSD, % |
|---|---|---|---|---|---|---|---|---|---|---|---|---|---|---|---|
| | Lab 1 | Lab 2 | Lab 3 | Lab 4 | Lab 5 | Lab 6 | Lab 7 | Lab 8 | Lab 9 | Lab 10 | Lab 11 | Lab 12 | Lab 13 | | |
| acetaldehyde | 1,114 | 1,532 | 1,584 | 1,403 | 1,506 | 1,636 | 1,524 | 1,672 | 1,596 | 1,622 | 1,301 | 1,627 | 1,648 | 1,52 | 10,6 |
| methyl acetate | 1,485 | 1,770 | 1,625 | 1,523 | 1,722 | 1,901 | 1,767 | 1,929 | 1,544 | 1,712 | 1,541 | 1,591 | 1,806 | 1,69 | 8,7 |
| ethyl acetate | 1,178 | 1,101 | 1,101 | 1,125 | 0,940 | 1,164 | 1,102 | 1,207 | 1,050 | 1,228 | 1,085 | 1,305 | 1,117 | 1,13 | 7,9 |
| methanol | 1,302 | 1,294 | 1,297 | 1,351 | 1,335 | 1,337 | 1,425 | 1,325 | 1,286 | 1,414 | 1,347 | 1,449 | 1,215 | 1,34 | 4,8 |
| 2-propanol | 0,972 | 0,953 | 1,002 | 0,968 | 0,916 | 0,927 | 0,921 | 0,975 | 0,997 | 0,803 | 0,861 | 0,962 | 0,955 | 0,94 | 5,9 |
| 1-propanol | 0,775 | 0,814 | 0,783 | 0,748 | 0,763 | 0,760 | 0,802 | 0,757 | 0,773 | 0,803 | 0,717 | 0,852 | 0,704 | 0,77 | 5,1 |
| isobutanol | 0,620 | 0,641 | 0,631 | 0,631 | 0,642 | 0,650 | 0,686 | 0,609 | 0,695 | 0,664 | 0,604 | 0,708 | 0,552 | 0,64 | 6,5 |
| 1-butanol | 0,646 | 0,655 | 0,657 | 0,673 | 0,690 | 0,705 | 0,699 | 0,685 | 0,760 | 0,718 | 0,640 | 0,772 | 0,610 | 0,69 | 6,8 |
| isoamylol | 0,581 | 0,595 | 0,599 | 0,613 | 0,656 | 0,648 | 0,686 | 0,616 | 0,671 | 0,660 | 0,607 | 0,715 | 0,586 | 0,63 | 6,6 |
| | response facotor (RF), pC | | | | | | | | | | | | | RF average, pC | RSD, % |
| ethanol | 144,67 | 85,21 | 91,39 | 172,88 | 102,35 | 67,36 | 142,48 | 79,23 | 42,47 | 79,37 | 8046,72 | 130,36 | 693,52 | 759,8 | 289,0 |

Values RF are varied in wide range from minimum value 42,47 pC from Lab 9 to maximum, value 693,52 pC from Lab 13. At the same time the relative standard deviations for values RRF experimentally obtained in different 13 laboratories do not exceed 10,6 % for acetaldehyde and do not exceed 8,7 % for all other analyzed volatile compounds. This fact allows us to use averaged values RRF for the primary estimating measurements of volatile compounds in alcohol products without graduation procedure of GC.

Thousands of analytical and testing laboratories all over the world carry out gas chromatographic analysis of volatile compounds in spirit drinks every day. Their employees may validate proposed new method in actual practice, making sure its simplicity, accessibility and effectiveness in everyday practice. The obtained results show the possibility of developing a new international standard of measurement procedure, which will allow increase the data accuracy and will considerably simplify the measurement procedure.

## 3. ACKNOWLEDGMENTS

We would like to thank the Winery and Distillery Plant "Chashniki" (Belarus) for supplying high-grade ethanol and New Analytical Systems Ltd. for instrumentation support.

**AUTHOR INFORMATION**

Corresponding Author
*Tel.: +375 296 51 33 91. E-mail: svcharapitsa@tut.by.